\documentclass{ws-ijmpa}

\begin{document}

\markboth{A.\,V.~Kuznetsov, N.\,V.~Mikheev, A.\,V.~Serghienko}
{The third type of fermion mixing in the lepton and quark interactions with leptoquarks}

%
\catchline{}{}{}{}{}
%

\title{THE THIRD TYPE OF FERMION MIXING IN THE LEPTON \\ AND QUARK INTERACTIONS WITH LEPTOQUARKS}


\author{\footnotesize A.\,V.~KUZNETSOV, N.\,V.~MIKHEEV, A.\,V.~SERGHIENKO}

\address{Division of Theoretical Physics, Department of Physics,\\
Yaroslavl State P.\,G.~Demidov University, Sovietskaya 14,\\
150000 Yaroslavl, Russian Federation\\
avkuzn@univ.uniyar.ac.ru, mikheev@univ.uniyar.ac.ru, serghienko@gmail.com}

\maketitle

\begin{history}
\received{Day Month Year}
\revised{Day Month Year}
\end{history}

\begin{abstract}
The low-energy manifestations of a minimal extension of the 
electroweak standard model based on the quark-lepton symmetry $SU(4)_V \otimes SU(2)_L
\otimes G_R$ of the Pati--Salam type are analyzed. Given this
symmetry the third type of mixing in the interactions of the
$SU(4)_V$ leptoquarks with quarks and leptons is shown to be
required. An additional arbitrariness of the mixing parameters could
allow, in principle, to decrease noticeably the lower bound on the
vector leptoquark mass originated from the low-energy rare
processes, strongly suppressed in the standard model.

\keywords{quark-lepton symmetry; vector leptoquark; fermion mixing}
\end{abstract}

\ccode{PACS numbers: 12.10.Dm, 13.20.-v, 14.80.Sv}

\newcommand{\ba}{\begin{aligned}}
\newcommand{\ea}{\end{aligned}}
\newcommand{\nn}{\nonumber}
\def\fc{\frac}
\def\hf{\fc{1}{2}}
\def\st{\sqrt}
\def\lm{\limits}
\def\df{\partial}
\def\prm{\prime}
\def\al{\alpha}
\def\ga{\gamma}
\def\de{\delta}
\def\si{\sigma}
\def\te{\theta}
\def\vp{\varphi}
\def\ep{\epsilon}
\def\ve{\varepsilon}
\def\vk{\varkappa}
\def\om{\omega}
\def\la{\lambda}
\def\l{\left}
\def\r{\right}
\def\td{\tilde}
\def\ls{\leqslant}
\def\pp{\perp}
\def\pr{\parindent}
\def\vs{\vspace}

\section{Introduction}
\label{sec:Introduction}

While the LHC methodically examines the energy scale of the electroweak theory and above, 
it is time to recall the two criteria for evaluating a physical theory, 
mentioned by A. Einstein.\cite{Einstein:1949}
The first point of view is obvious: a theory must not contradict empirical facts, 
and it is called the ``external confirmation''. 
The test of this criterion both for the standard model and its various extensions is now engaged in the LHC.  
The second point of view called the ``inner perfection'' of the theory, 
may be very important to refine the search area for new physics.

All existing experimental data in particle physics are in good agreement with 
the standard model predictions. However, the problems exist which could 
not be resolved within the standard model and it is obviously not a complete or final 
theory. It is unquestionable that the standard model should be the low-energy limit of 
some higher symmetry. The question is what could be this symmetry. And 
the main question is, what is the mass scale of this symmetry restoration. 
A gloomy prospect is the restoration of this higher symmetry at once on a 
very high mass scale, the so-called gauge desert. A concept of a consecutive 
symmetry restoration is much more attractive. It looks natural in this case to 
suppose a correspondence of the hierarchy of symmetries and the hierarchy 
of the mass scales of their restoration. Now we are on the first step of some 
stairway of symmetries and we try to guess what could be the next one. If 
we consider some well-known higher symmetries from this point of view, two 
questions are pertinent. First, isn't the supersymmetry\cite{Nilles:1984} as the symmetry of 
bosons and fermions, higher than the symmetry within the fermion sector, 
namely, the quark-lepton symmetry,\cite{Pati:1974} or the symmetry within the boson 
sector, namely, the left-right 
symmetry?\cite{Lipmanov:1967}\cdash\cite{Beg:1977} 
Second, wouldn't the supersymmetry 
restoration be connected with a higher mass scale than the others? 
The recent searches for supersymmetry carried out
at the Tevatron and the LHC colliders\cite{Portell_Bueso:2011} 
shown that no significant deviations from the
standard model predictions have been found, the vast parameter
space available for supersymmetry has been substantially reduced
and the most probable scenarios predicted by
electroweak precision tests are now excluded or under
some constraints after the new stringent limits.

We should like to analyse a possibility when the quark-lepton symmetry 
is the next step beyond the standard model. 
Along with the ``inner perfection'' argument for this theory, 
there exists a direct evidence in favor of it. 
The puzzle of fermion generations is recognized as one of the most 
outstanding problems of present particle physics, and may be the main justification for the need 
to go beyond the standard model. Namely, the cancellation of triangle axial anomalies which is 
necessary for the standard model to be renormalized, requires that fermions be grouped into generations.
This association provides an equation $\sum_f \, T_{3f} \, Q_f^2 = 0$, where the summation is taken over all 
fermions of a generation, both quarks of three colors and leptons, 
$T_{3f}$ is the 3d component of the weak isospin, and $Q_f$ is the electric charge of a fermion. 
Due to this equation, the divergent axial-vector part of the triangle $Z \gamma \gamma$ diagram 
with a fermion loop vanishes. 

The model where a combination of quarks and leptons into generations looked the most natural, 
proposed by J.C. Pati and A. Salam\cite{Pati:1974} was based on the 
quark-lepton symmetry. The lepton number was treated in the model as the fourth color. 
As the minimal gauge group realizing this symmetry, one can consider the semi-simple group $SU(4)_V \otimes SU(2)_L \otimes G_R$. To begin with, one can take the group $U(1)_R$ as $G_R$. 
The fermions were combined into the fundamental representations of the 
$SU(4)_V$ subgroup, the neutrinos with the \emph{up} quarks and the charged leptons 
with the \emph{down} quarks: 
\begin{equation}
\left ( \begin{array}{c} u^1 \\ u^2 \\ u^3 \\ \nu \end{array}
\right )_i \, , \qquad \left (
\begin{array}{c} d^1 \\ d^2 \\ d^3 \\ \ell \end{array} \right )_i \, ,
\qquad i=1,2,3 \dots \, (?) \,, 
\label{eq:q}
\end{equation}
where the superscripts 1,2,3 number colors and the subscript $i$ numbers fermion generations, 
i.e. $u_i$ denotes $u, c, t, \dots$ and $d_i$ denotes $d, s, b, \dots$.

The left-handed fermions form fundamental representations of the
$SU(2)_L$ subgroup:
\begin{equation}
\left ( \begin{array}{c} u^c \\ d^c \end{array}
\right )_L \, , \qquad 
\left ( \begin{array}{c} \nu \\ \ell \end{array} \right )_L \, .
\label{eq:d}
\end{equation}
One should keep in mind that  
when considering the mass eigenstates, it is necessary to take into 
account the mixing of fermion states~\eqref{eq:q}, \eqref{eq:d}, 
to be analysed below.  
 
Let us remind that such an extension of
the standard model has a number of attractive features. 

\begin{enumerate}
\item
As it was mentioned above, definite quark-lepton symmetry is necessary in order that the
standard model be renormalized: cancellation of triangle anomalies
requires that fermions be grouped into generations.

\item
There is no proton decay because the lepton charge treated as the
fourth color is strictly conserved.

\item
Rigid assignment of quarks and leptons to representations~\eqref{eq:q} leads to 
a natural explanation for a fractional quark hypercharge. Indeed, 
the traceless 15-th generator $T_{15}^V$ of the $SU(4)_V$ subgroup can be represented in
the form 
\begin{equation} 
T_{15}^V=\st{\fc{3}{8}} \; \text{diag}\l(\fc{1}{3}\,,\,\fc{1}{3}\,,\,\fc{1}{3}\,,\,-1\r) 
=\st{\fc{3}{8}} \; Y_V \,. 
\label{eq:T15}
\end{equation} 
It is remarkable that the values of the standard model hypercharge of the left-handed quarks and leptons 
combined into the $SU(2)_L$ doublets turn out to be placed on the diagonal. 
Let us call it the vector hypercharge, $Y_V$, and assume that it belongs to both 
the left- and right-handed fermions. 

\item
Let us suppose that $G_R = U(1)_R$. The well-known values 
of the standard model hypercharge of the left and right, 
and \emph{up} and \emph{down} quarks and 
leptons are:
\begin{equation}
Y_{SM} \; = \; \left \{
\begin{array}{c} \left (\begin{array}{c} \frac{1}{3} \\ \\ \frac{1}{3}
\end{array} \right ) \quad \mbox{for} \; q_L ; \\ \\
\left (\begin{array}{c} \frac{4}{3} \\ \\ -\frac{2}{3}
\end{array} \right ) \quad \mbox{for} \; q_R ; \end{array}
\begin{array}{c} \left (\begin{array}{c} - 1 \\ \\ - 1
\end{array} \right ) \quad \mbox{for} \; \ell_L \\ \\
\left (\begin{array}{c} 0 \\ \\ - 2
\end{array} \right ) \quad \mbox{for} \; \ell_R \end{array}
\right \}.
\label{eq:Y}
\end{equation}
Then, from the equation $Y_{SM} = Y_V + Y_R$, taking Eq.~\eqref{eq:T15} into account, 
one obtains that the values of the right 
hypercharge $Y_R$ occur to be equal $\pm 1$ for the \emph{up} and \emph{down} 
fermions correspondingly, both quarks and leptons. 
It is tempting to interpret this circumstance as the indication
that the right-hand hypercharge is the doubled third component of
the right-hand isospin. Thus, the subgroup $G_R$ may be $SU(2)_R$.
\end{enumerate}

``Under these circumstances one would be surprised if Nature had made no use of it'', 
as P.~Dirac wrote on another occasion.\cite{Dirac:1931} 

The most exotic object of the Pati--Salam type symmetry is the charged 
and colored gauge $X$ boson named leptoquark. Its mass $M_X$ should be the 
scale of breaking of $SU(4)_V$ to $SU(3)_c$. 
Bounds on the vector leptoquark mass are obtained both directly and
indirectly, see Ref.~\refcite{Nakamura:2010}, pp. 490-494, and Ref.~\refcite{PDG:2011}.
The recent direct search\cite{Aaltonen:2008_LQ} for vector leptoquarks using $\tau^+ \tau^- b \bar b$ 
events in $p \bar p$ collisions at $E_{cm} = 1.96$ TeV have provided the lower mass limit at a level 
of 250--300 GeV, depending on the coupling assumed.  
Much more stringent indirect limits are calculated from the bounds
on the leptoquark-induced four-fermion interactions, which are
obtained from low-energy experiments. There is an extensive series of papers where 
such indirect limits on the vector leptoquark mass were estimated, 
see e.g. Refs.~\refcite{Shanker:1982}--\refcite{Smirnov:2008}. 
The most stringent bounds\cite{Nakamura:2010} 
were obtained from the data on the $\pi \to e \nu$ decay and 
from the upper limits on the $K_L^0 \to e \mu$ and $B^0 \to e \tau$ decays. 
However, those estimations were not comprehensive 
because the phenomenon of a mixing in the lepton-quark currents was not 
considered there. It will be shown that such a mixing inevitably occurs in 
the theory. 

An important part of the model under consideration is its scalar sector, which 
also contains exotic objects such as scalar leptoquarks. We do not concern here 
the scalar sector, which could be much more ambiguous than the gauge one. Such an analysis can be 
found e.g. in Refs.~\refcite{Smirnov:2007,Smirnov:2008,Leurer:1994a}.

The paper is organized as follows. 
In Sec.~\ref{sec:Mixing}, it is argued that three types of fermion mixing 
inevitably arise at the loop level if initially fermions are taken without mixing. 
The effective four-fermion Lagrangian caused by the leptoquark interactions with 
quarks and leptons is presented in Sec.~\ref{sec:Lagrangian}. 
In Sec.~\ref{sec:Constraints}, the constraints on the parameters of the scheme 
are obtained based on the data from different low-energy processes which are 
strongly suppressed or forbidden in the standard model. 
Combined constraint on the vector leptoquark mass from the $\pi$, $K$, $\tau$ and $B$ decays
is obtained in Sec.~\ref{sec:Combination}.  
In Sec.~\ref{sec:Different}, combined constraint on the vector leptoquark mass from the same 
processes is obtained in the case of different mixings for left-handed and right-handed fermions. 
In Sec.~\ref{sec:Pi_decay}, one more mixing independent bound on the vector leptoquark mass 
is presented, coming from the decay $\pi^0\to\nu\bar{\nu}$. 

\section{The third type of fermion mixing}
\label{sec:Mixing}

As the result of the Higgs mechanism in the Pati--Salam model,
fractionally charged colored gauge $X$-bosons, vector leptoquarks appear. 
Leptoquarks are responsible for transitions between
quarks and leptons. The scale of the breakdown of $SU(4)_V$ symmetry
to $SU(3)_c$ is the leptoquark mass $M_X$. The three fermion
generations are grouped into the following $\{4,2\}$ representations
of the $SU(4)_V\otimes SU(2)_L$ group:
\begin{equation}
\begin{pmatrix} u^c & d^c\\
\nu & \ell
\end{pmatrix}_i~\l(i=1,2,3\r).
\label{eq:mixing}
\end{equation}
where $c$ is the color index to be further omitted. 
It is known that there exists the mixing of quarks
in weak charged currents, which is described by the
Cabibbo-Kobayashi-Maskawa matrix. Therefore, at least one of the
states in \eqref{eq:mixing}, $u$ or $d$, is not diagonal in mass. It can easily be seen
that, because of mixing that arises at the loop level, none of the
components is generally a mass eigenstate. As usual, we assume that
all the states in \eqref{eq:mixing}, with the exception of $d$, are initially diagonal in
mass. This leads to nondiagonal transitions $\ell \to X + d (s,b) \to \ell^\prm$
through a quark-leptoquark loop, see Fig.~1. As this diagram is divergent, the corresponding 
counterterm should exist at the tree level. This means that the lepton states $\ell$ 
in \eqref{eq:mixing} are not the mass eigenstates, and there is mixing in the
lepton sector. Other nondiagonal transitions arise in a similar way.
Hence, in order that the theory be renormalizable, it is necessary
to introduce all kinds of mixing even at the tree level.
%
\begin{figure}
\begin{center}
\includegraphics*[width=0.3\textwidth]{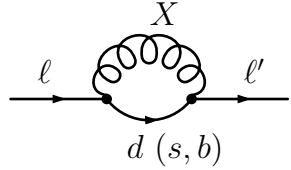}
\caption{Feynman diagram illustrating the appearance of fermion mixings.} 
\end{center}
\label{fig:X_loop}
\end{figure}
%
As all the fermion representations are identical, they can always be
regrouped in such a way that one state is diagonal in mass. The most
natural way is to diagonalize charged leptons. In this case, fermion
representations can be written in the form
\begin{equation} 
\begin{pmatrix} u & d\\
\nu & \ell
\end{pmatrix}_{\ell}=
\begin{pmatrix} u_e & d_e\\
\nu_e & e
\end{pmatrix},~
\begin{pmatrix} u_\mu & d_\mu\\
\nu_\mu & \mu
\end{pmatrix},~
\begin{pmatrix} u_\tau & d_\tau\\
\nu_\tau & \tau
\end{pmatrix}.
\label{eq:repr1} 
\end{equation}
Here, the quarks and neutrinos subscripts $\ell=e,\mu,\tau$ 
label the states which are not mass eigenstates and which enter into the
same representation as the charged lepton $\ell$: 
\begin{equation} 
\nu_{\ell}=\sum_i {\cal K}_{{\ell} i}\nu_i \,, \quad 
u_{\ell}=\sum_p {\cal U}_{{\ell} p}u_p\,, \quad 
d_{\ell}=\sum_n {\cal D}_{{\ell} n}d_n \,. 
\label{eq:repr2} 
\end{equation}
Here, ${\cal K}_{{\ell}i}$ is the unitary leptonic mixing matrix 
by Pontecorvo--Maki--Nakagawa--Sakata.\cite{Pontecorvo:1957}\cdash\cite{Maki:1962}
The matrices ${\cal U}_{{\ell}p}$ and ${\cal D}_{{\ell}n}$ are the
unitary mixing matrices in the interactions of leptoquarks with the \emph{up} and \emph{down} 
fermions correspondingly, both quarks and leptons. 
The states $\nu_i,~u_p$ and $d_n$ are the mass eigenstates: 
\begin{equation} \ba &\nu_i=\l(\nu_1,\nu_2,\nu_3\r),\\
&u_p=\l(u_1,u_2,u_3\r)=\l(u,c,t\r),\\
&d_n=\l(d_1,d_2,d_3\r)=\l(d,s,b\r). \ea 
\label{eq:repr3} 
\end{equation} 
Thus, there are generally three types of mixing in this scheme.

In our notation, the well-known Lagrangian describing the
interaction of charge weak currents with $W$-bosons takes the form
\begin{eqnarray} 
{\cal L}_W &=& \fc{g}{2\st{2}}\l[\l(\bar{\nu}_\ell O_\al \ell \r)+\l(\bar{u}_\ell O_\al d_\ell \r)\r] W^*_\al + \text{h.c.}
\nonumber\\
&=&\fc{g}{2\st{2}}\l[{\cal K}^*_{\ell i}\l(\bar{\nu}_i O_\al \ell \r) 
+ {\cal U}^*_{\ell p} {\cal D}_{\ell n} \l(\bar{u}_p O_\al d_n \r)\r]W^*_\al + \text{h.c.},
\label{eq:Lagr_W} 
\end{eqnarray} 
where $g$ is the constant of the $SU(2)_L$ group and
$O_\al=\ga_\al\l(1-\ga_5\r)$. It follows that the standard
Cabibbo--Kobayashi--Maskawa matrix is $V={\cal U}^\dagger {\cal D}$. This is
the only available information about the matrices ${\cal U}$ and
${\cal D}$ of mixing in the leptoquark sector. The matrix ${\cal K}$
describing a mixing in the lepton sector is the subject of intensive
experimental studies.

Following the spontaneous breakdown of the $SU(4)_V$ symmetry to $SU(3)_c$
on the scale of $M_X$, six massive vector bosons forming three charged
colored leptoquarks, decouple from the 15-plet of gauge fields. The
interaction of these leptoquarks with fermions has the form 
\begin{equation}
{\cal L}_X=\fc{g_S\l(M_X\r)}{\st{2}} \l[ {\cal D}_{\ell n} \l(\bar{\ell} \ga_\al d^c_n\r) + 
\l({\cal K}^\dagger {\cal U}\r)_{ip} \l(\bar{\nu}_i \ga_\al u^c_p \r) \r] X^c_\al + \text{h.c.},
\label{LagDen} 
\end{equation} 
where the color superscript $c$ is written
explicitly once again. The coupling constant $g_S\l(M_X\r)$ is
expressed in terms of the strong-interaction constant $\al_S$ on the
scale of the leptoquark mass $M_X$ as
$g_S^2\l(M_X\r)/4\pi=\al_S\l(M_X\r)$.

\section{Effective Lagrangian with allowance for QCD corrections}
\label{sec:Lagrangian}

If the momentum transfer satisfies the condition $q^2\ll M_X^2$,
the Lagrangian \eqref{LagDen} leads to the effective four-fermion
vector-vector interaction between quarks and leptons. By applying
the Fierz transformation, we can isolate the lepton and quark currents
(scalar, pseudoscalar, vector and axial-vector currents) in the
effective Lagrangian. In constructing the effective Lagrangian of
leptoquark interactions, it is necessary to take into account the QCD
corrections, which can easily be estimated, see e.g. Refs.~\refcite{Vainstein,Vysotskii}. 
In the case under study, we
can use the approximation of leading logarithms because
$\ln\l(M_X/\mu\r)\gg1$, where $\mu\sim1~\text{GeV}$ is the typical
hadronic scale. As the result of taking the QCD corrections into
account, the scalar and pseudoscalar coupling constants acquire the
enhancement factor 
\begin{equation}
Q\l(\mu\r) = \l(\fc{\al_S\l(\mu\r)}{\al_S\l(M_X\r)}\r)^{4/\bar{b}},
\label{EnFact} 
\end{equation} 
where $\al_S\l(\mu\r)$ is the strong-interaction
constant on the scale $\mu$, $\bar{b}=11-2/3\l(\bar{n}_f\r)$, and
$\bar{n}_f$ is the mean number of quark flavors on the scales
$\mu^2\leq q^2\leq M_X^2$; for $M_X^2\gg m_t^2$, we have
$\bar{b}\simeq7$.

Let us investigate the contribution to low-energy processes from the 
interaction Lagrangian \eqref{LagDen} involving leptoquarks and find
constraints on the parameters of the scheme from available
experimental data. It will be shown below that the most stringent
constraints on the vector-leptoquark mass $M_X$ and on the elements
of the mixing matrix ${\cal D}$ follow from the data on rare 
$\pi$ and $K$ decays and on $\mu e$ conversion on nuclei. 

Possible constraints on the masses and coupling constants of vector
leptoquarks from experimental data on rare $\pi$ and $K$ meson decays were
analyzed in Refs.~\refcite{Shanker:1982}--\refcite{Smirnov:2008}. 
One approach\cite{Shanker:1982,Leurer:1994b,Davidson:1994} was based on using 
the phenomenological model-independent Lagrangians describing the interactions of
leptoquarks with quarks and leptons. 
Pati--Salam quark-lepton symmetry was considered in
Refs.~\refcite{Deshpande:1983,Valencia:1994}--\refcite{Smirnov:2008}. 
QCD corrections were included into an analysis in Refs.~\refcite{Valencia:1994}--\refcite{Kuznetsov:1995}. 
The authors of Ref.~\refcite{Valencia:1994} considered
the possibility of mixing in quark-lepton currents, but they
analyzed only specific cases in which each charged lepton is
associated with one quark generation. In our notation, this
corresponds to the matrices ${\cal D}$ that are obtained from the
unit matrix by making all possible permutation of columns.

In the description of the $\pi$- and $K$-meson interactions, it is
sufficient to retain only the scalar and pseudoscalar coupling
constants in the effective Lagrangian. Really, these couplings are more
significant in the amplitudes, because they are enhanced, first, by QCD corrections, 
and second, by the smallness of the current-quark masses arising in 
the amplitude denominators. 
The corresponding part of the effective Lagrangian can be represented as 
\begin{eqnarray} 
\Delta{\cal L}_{\pi,K} &=& -\fc{2\pi\al_S\l(M_X\r)}{M_X^2} \, Q \l(\mu\r) 
\l [{\cal D}_{\ell n} \l( {\cal U}^\dagger {\cal K}\r)_{pi} 
\l(\bar{\ell} \ga_5 \nu_i \r) \l(\bar{u}_p\ga_5d_n\r) + \text{h.c.} 
- \l(\ga_5\to1\r)\r]
\nonumber\\
&-&\fc{2\pi\al_S\l(M_X\r)}{M_X^2} \, Q \l(\mu\r) 
\bigg[{\cal D}_{\ell n} {\cal D}^*_{\ell^\prm n^\prm} 
\l(\bar{\ell} \ga_5 \ell^\prm \r) \l( \bar{d}_{n^\prm} \ga_5 d_n \r)  
\nonumber\\
&+& \l({\cal K}^\dagger {\cal U}\r)_{ip}\l({\cal U}^\dagger {\cal K}\r)_{p^\prm
i^\prm}\l(\bar{\nu}_i\ga_5\nu_{i^\prm}\r)\l(\bar{u}_{p^\prm}\ga_5u_p\r)-\l(\ga_5\to1\r)
\bigg].
\label{LagPiK}  
\end{eqnarray} 
This Lagrangian contributes to the rare $\pi$, $K$, $\tau$ and $B$ decays, 
which are strongly suppressed or forbidden in the standard model.

\section{Constraints on the parameters of the scheme from low-energy processes}
\label{sec:Constraints}

\subsection{$\mu e$ Universality in $\pi_{\ell 2}$ and $K_{\ell 2}$ Decays}

\begin{figure}
\begin{center}
\includegraphics*[width=0.67\textwidth]{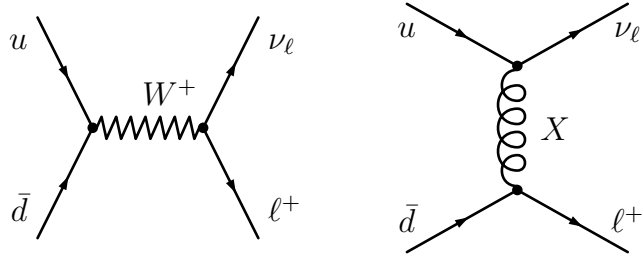}
\caption{Feynman diagrams for the $\pi^+ (u \bar{d}) \to \ell^+ \nu_{\ell}$ decay via the $W$-boson 
and leptoquark $X$ exchange. Substituting the $c$ quark instead of $u$ and 
other \emph{down} antiquarks $\bar{s}, \bar{b}$ instead of $\bar{d}$, 
one obtains the diagrams for lepton decays 
of various charged mesons, where the $W$ and $X$ boson contributions interfere.} 
\end{center}
\label{fig:pienu}
\end{figure}

Analysis reveals that, in contrast to $W$-boson contribution, the
leptoquark contribution to the decay $\pi\to e\nu$, see Fig.~2, does not involve
suppression that is due to the electron mass. The corresponding
part of the amplitude can be represented in the form 
\begin{equation}
\Delta{\cal M}^X_{\pi e\nu} = - \fc{2\pi\al_S\l(M_X\r)}{M_X^2}{\cal
D}_{ed}\l({\cal U}^\dagger {\cal K}\r)_{ui}\fc{f_\pi
m_\pi^2Q\l(\mu\r)}{m_u\l(\mu\r)+m_d\l(\mu\r)}\l(\bar{e}\ga_5\nu_i\r),
\label{Amp1} 
\end{equation} 
where $f_\pi\simeq132~\text{MeV}$ is the constant of the
$\pi l\nu$ decay, and $m_{u,d}\l(\mu\r)$ are the running quark
masses on the scale $\mu$. The ratio $Q\l(\mu\r)/m\l(\mu\r)$ is a
renormalization-group invariant because the function $Q\l(\mu\r)$
\eqref{EnFact} determines the law of variation of the running mass. The
known values\cite{Nakamura:2010} of the current-quark masses, $m_u = (1.7-3.3)~\text{MeV}$,
$m_d = (4.1-5.8)~\text{MeV}$, and $m_s = 101^{+29}_{-21}~\text{MeV}$, correspond to
the scale $\mu\simeq 2~\text{GeV}$. For evaluations, we take the central values of them. 
The contribution to the amplitude from the 
$W$-boson exchange has the form 
\begin{equation} 
\Delta{\cal M}^W_{\pi e\nu} = - f_\pi\fc{G_F}{\st{2}}m_e{\cal
K}_{ei}V_{ud}\l[\bar{e}\l(1-\ga_5\r)\nu_i\r]. 
\label{Amp2} 
\end{equation}

Taking into account the interference of the amplitudes \eqref{Amp1} and
\eqref{Amp2}, we find that the ratio $\Gamma\l(\pi\to
e\nu\r)/\Gamma\l(\pi\to\mu\nu\r)\equiv R_\pi$ of the decay widths is
given by 
\begin{equation} 
R_\pi=R_{\pi W}\l[1-\fc{2\st{2}\pi\al_S\l(M_X\r)m_\pi^2Q}{G_FM_X^2m_e\l(m_u+m_d\r)} 
\text{Re}\l(\fc{{\cal D}_{ed}{\cal U}^*_{eu}}{V_{ud}}\r)\r], 
\label{RatioPi} 
\end{equation} 
where
$R_{\pi W}=\l(1.2345\pm0.0010\r)\times10^{-4}$ is the value of this
ratio in the standard model.\cite{Marciano:1993} Bringing together the
results of $R_\pi$ measurements at TRIUMF\cite{Britton:1992a,Britton:1992b} and in the Paul Scherrer
Institute,\cite{Czapek:1993} we arrive at the conclusion
that the leptoquark mass obeys the constraint
\begin{equation} 
M_X>\l(210~\text{TeV}\r)\l|\text{Re}
\l({\cal D}_{ed}{\cal U}^*_{eu}\r)\r|^{1/2}. 
\label{Con1} 
\end{equation} 

The vector-leptoquark contribution can also disturb the ratio $R_K$
of $\mu e$ universality for the decays $K\to e\nu$ and $K\to\mu\nu$.
Feynman diagrams for these processes can be obtained from the diagrams in Fig.~2, 
replacing $\bar d$ by $\bar s$. 
In an analogy with the analysis of the ratio of
$\pi_{\ell 2}$-decay widths, we find that the experimental data\cite{AMBROSINO09} on $R_K$
 yield the constraint 
\begin{equation}
M_X>\l(150~\text{TeV}\r) \l|\text{Re} 
\l({\cal D}_{es}{\cal U}^*_{eu}\r)\r|^{1/2}. 
\label{Con2} 
\end{equation} 
In this analysis, it was essential to take into account the
interference of the $W$-boson and leptoquark contributions to the
amplitudes, which was not considered in Ref.~\refcite{Valencia:1994}.  
Really, because of mixing,
the phenomenological neutrino $\nu_u$ produced in the leptoquark
interaction is a superposition of the phenomenological neutrinos
produced in the weak interaction: 
\begin{equation} 
\nu_u={\cal U}^*_{eu}\nu_e+{\cal U}^*_{\mu u}\nu_\mu+{\cal U}^*_{\tau u}\nu_{\tau}. 
\label{PhenNeut1}
\end{equation} 
Because the experimental interval for $R_K$
(see Table~\ref{tab1}) falls below the theoretical value of this ratio in the
standard model ($R_{KW}=2.57\times10^{-5}$) and because the
interference of the above amplitudes is destructive, the inclusion
of this interference substantially changes the estimate of the
leptoquark mass.

\subsection{Rare $K_L^0$-Meson Decays}

\begin{figure}
\begin{center}
\includegraphics*[width=0.67\textwidth]{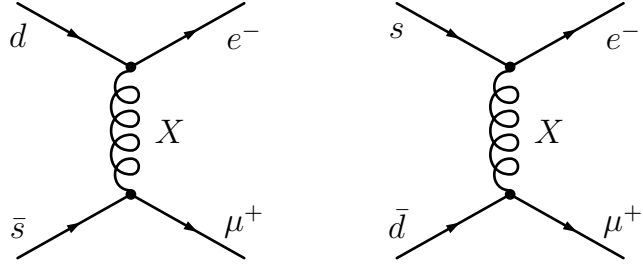}
\caption{Feynman diagrams for the $K^0_L (d \bar{s}+s \bar{d}) \to e^- \mu^+$ decay 
forbidden in the standard model, via the leptoquark exchange.} 
\end{center}
\label{fig:K0emu}
\end{figure}

The amplitude of the process $K_L^0\to e^-\mu^+$ forbidden in the standard model, is calculated in
the similar way as the amplitude \eqref{Amp1}, see Fig.~3. The result is 
\begin{equation} 
{\cal M}^X_{Ke\mu}=\fc{\st{2}\pi\al_S\l(M_X\r)f_Km_K^2Q}{M_X^2\l(m_s+m_d\r)}
\l({\cal D}_{ed}{\cal D}^*_{\mu s}+{\cal D}_{es}{\cal D}^*_{\mu
d}\r)\l(\bar{e}\ga_5\mu\r), 
\label{Amp3} 
\end{equation} 
where $f_K\simeq160~\text{MeV}$ is the constant of the $Kl\nu$ decay.
We find that the use of the available
experimental data\cite{AMBROSE98B} in our scheme leads to the
constraint 
\begin{equation} 
M_X>\l(2100~\text{TeV}\r) 
\l|{\cal D}_{ed}{\cal D}^*_{\mu s}+{\cal D}_{es}{\cal D}^*_{\mu d}\r|^{1/2}. 
\label{Con3} 
\end{equation}

Experimental values\cite{AMBROSE00} of $\text{Br}\l(K_L^0\to\mu^+\mu^-\r)$ closely
approach the unitary limit $\text{Br}_{\text{abs}}=6.8\times10^{-9}$.
Therefore, the effective leptoquark contribution to
$\text{Br}\l(K_L^0\to\mu^+\mu^-\r)$ is unlikely to exceed
$1\times10^{-10}$. The amplitude of the process is obtained
from \eqref{Amp3} by making the substitution $e\to\mu$. We finally
obtain 
\begin{equation} 
M_X>\l(1100~\text{TeV}\r)\l| \text{Re}\l({\cal D}_{\mu d}{\cal D}^*_{\mu s}\r)\r|^{1/2}. 
\label{Con4} 
\end{equation}

The amplitude of one more rare $K_L^0$ decay, into an
electron and a positron through an intermediate leptoquark, can
also be obtained from \eqref{Amp3} by means of the substitution $\mu\to e$. 
Experimental values\cite{AMBROSE98} of $\text{Br}\l(K_L^0\to e^+ e^-\r)$ closely
approach the unitary limit $\text{Br}_{\text{abs}}=9\times10^{-12}$.
Therefore, the effective leptoquark contribution to
$\text{Br}\l(K_L^0\to e^+ e^- \r)$ is unlikely to exceed
$5\times10^{-12}$. In this case, the constraint on the leptoquark mass is 
\begin{equation}
M_X>\l(2400~\text{TeV}\r)\l|\text{Re}\l({\cal D}_{ed}{\cal D}^*_{es}\r)\r|^{1/2}. 
\label{Con5} 
\end{equation}

\begin{table}[t]
\tbl{Constraints on the leptoquark masses and on the elements of
the mixing matrices from experimental data on rare $\pi$ and $K$
decays and on $\mu e$ conversion on nuclei.}
{\begin{tabular}{lcl} \toprule
Experimental limit & Ref. & Bound \\ \colrule 
\bigskip
$\frac{\mbox{\footnotesize $\Gamma\left(\pi^+ \to e^+ \nu_e\right)$}}
{\mbox{\footnotesize $\Gamma\left(\pi^+ \to \mu^+ \nu_\mu\right)$}} 
= (1.2310\pm0.0037) \times 10^{-4}$ 
& \refcite{Britton:1992a}--\refcite{Czapek:1993} & $\frac{\mbox{\footnotesize
$M_X$}} {\mbox{\footnotesize $|\text{Re}({\cal D}_{ed} {\cal U}^*_{eu})|^{1/2}$}} \,
> \, 210~\textrm{TeV}$ \\
\bigskip
$\frac{\mbox{\footnotesize $\Gamma\left(K^+ \to e^+ \nu_e\right)$}}
{\mbox{\footnotesize $\Gamma\left(K^+ \to \mu^+ \nu_\mu\right)$}} 
= (2.493\pm0.031) \times 10^{-5}$ 
& \refcite{AMBROSINO09} & $\frac{\mbox{\footnotesize $M_X$}}
{\mbox{\footnotesize $|\text{Re}({\cal D}_{es} {\cal U}^*_{eu})|^{1/2}$}} \,
> \, 150~\textrm{TeV}$ \\
\bigskip
$Br(K^+ \to \pi^+ \mu^+ e^-) < 1.3 \times 10^{-11}$ 
& \refcite{SHER05} & $\frac{\mbox{\footnotesize $M_X$}}
{\mbox{\footnotesize $|{\cal D}_{ed} {\cal D}_{\mu s}|^{1/2}$}} \,
> \, 240~\textrm{TeV}$ \\
\bigskip
$Br(K^+ \to \pi^+ \mu^- e^+) < 5.2 \times 10^{-10}$ 
& \refcite{APPEL00} & $\frac{\mbox{\footnotesize $M_X$}}
{\mbox{\footnotesize $|{\cal D}_{es} {\cal D}_{\mu d}|^{1/2}$}} \,
> \, 100~\textrm{TeV}$ \\
\bigskip
$Br(K^0_L \to \mu^+ \mu^-) = (6.84\pm0.11) \times 10^{-9}$ 
& \refcite{AMBROSE00}, \refcite{ALEXOPOULOS04} & $\frac{\mbox{\footnotesize
$M_X$}} {\mbox{\footnotesize $|\text{Re}({\cal D}_{\mu d} {\cal D}^*_{\mu
s})|^{1/2}$}} \,
> \, 1100~\textrm{TeV}$ \\
\bigskip
$Br(K^0_L \to e^\pm \mu^\mp) < 4.7 \times 10^{-12}$ 
& \refcite{AMBROSE98B} & $\frac{\mbox{\footnotesize $M_X$}}
{\mbox{\footnotesize $|{\cal D}_{ed} {\cal D}^*_{\mu s}+{\cal D}_{es}
{\cal D}^*_{\mu d}|^{1/2}$}} \,
> \, 2100~\textrm{TeV}$ \\
\bigskip
$Br(K_L^0 \to e^+ e^-) = \left(9^{+6}_{-4}\right) \times 10^{-12}$ 
& \refcite{AMBROSE98} & $\frac{\mbox{\footnotesize $M_X$}}
{\mbox{\footnotesize $|\text{Re}({\cal D}_{ed} {\cal D}^*_{es})|^{1/2}$}} \,
> \, 2400~\textrm{TeV}$ \\
\bigskip
$\frac{\mbox{\footnotesize $\sigma\left(\mu^- \textrm{Au}\to e^- \textrm{Au}\right)$}}
{\mbox{\footnotesize $\sigma\left(\mu^- \textrm{Au}\to \textrm{capture}\right)$}} 
< 0.7 \times 10^{-12}$ 
& \refcite{Bertl:2006} & $\frac{\mbox{\footnotesize $M_X$}}
{\mbox{\footnotesize $|{\cal D}_{ed} {\cal D}_{\mu d}|^{1/2}$}} \,
> \, 1000~\textrm{TeV}$ \\
\botrule
\end{tabular}
\label{tab1}}
\end{table}

\subsection{Rare $K^+$ Decays}

Among rare $K^+$ decays, that can occur at the tree level in the model
under study, $K^+\to\pi^+\mu^+e^-$ \cite{SHER05} and
$K^+\to\pi^+\mu^-e^+$ \cite{APPEL00} yield the most stringent
constraints. The amplitude of the decay $K^+\to\pi^+\mu^+e^-$ can be
represented in the form 
\begin{equation} 
{\cal M}^X_{K\pi\mu e} 
= -\fc{2\pi\al_S\l(M_X\r)}{M_X^2}\fc{f^0_+\l(q^2\r)\l(m_K^2-m_\pi^2\r)+f^0_-\l(q^2\r)q^2}
{m_s-m_d} \, Q{\cal D}_{ed}{\cal D}^*_{\mu s}\l(\bar{e}\mu\r), 
\label{Amp4}
\end{equation} 
where $q$ is the 4-momentum of the lepton pair, and $f^0_{+,-}$
are the known form factors of the $K^0_{\ell 3}$ decay. The amplitude of the
decay $K^+\to\pi^+\mu^-e^+$ is obtained from \eqref{Amp4} by means of
the substitution $e\leftrightarrow\mu$. The resulting constraints on
the leptoquark mass involve the same elements as those present in
\eqref{Con3}, but they appear separately: 
\begin{equation} 
\ba
&M_X>\l(240~\text{TeV}\r)\l|{\cal D}_{ed}{\cal D}_{\mu
s}\r|^{1/2},\\
&M_X>\l(100~\text{TeV}\r)\l|{\cal D}_{es}{\cal D}_{\mu d}\r|^{1/2}.
\label{Con6} 
\ea 
\end{equation}

\subsection{$\mu e$ Conversion on a Nucleus}

This is one more low-energy process, which can
proceed through leptoquarks. Coherent $\mu e$ conversion, which
leaves the nucleus in the ground state and which leads to the
production of monoenergetic electrons with highest possible energy
$\simeq m_\mu$, is most convenient for observation. The effective
Lagrangian of the coherent $\mu e$ conversion involves only scalar and
vector quark currents. In the model under study, it has the form 
\begin{equation}
\Delta{\cal L}_{\mu e} 
=-\fc{2\pi\al_S \l(M_X\r)}{M_X^2} \, {\cal D}_{ed}{\cal D}^*_{\mu d} 
\l[\hf\l(\bar{e}\ga_\al\mu\r)\l(\bar{d}\ga_\al d\r) 
- \l(\bar{e}\mu\r)\l(\bar{d}d\r)Q\l(\mu\r)\r]. 
\label{LagNuc} 
\end{equation}
Using the computational technique developed in Ref.~\refcite{Shanker:1979} for
the effective quark-lepton interaction of the form \eqref{LagNuc}, we
estimated the branching ratio of $\mu e$ conversion on gold. By
applying the result to the experimental data reported in
Ref.~\refcite{Bertl:2006}, we arrived at the constraint 
\begin{equation}
M_X>\l(1000~\text{TeV}\r)\l|{\cal D}_{ed}{\cal D}_{\mu d}\r|^{1/2}.
\label{Con7} 
\end{equation}

The constraints that we obtained for the parameters of our model 
from experimental data on rare $\pi$ and $K$ decays and on $\mu e$ conversion 
on nuclei are summarized in Table~\ref{tab1}. It can be seen that all the constraints
involve the elements of the unknown unitary mixing matrices 
${\cal D}$ and ${\cal U}$, which are related by the single condition 
${\cal U}^+{\cal D}=V$. Therefore, the possibility cannot be ruled out 
for the constraints on the vector leptoquark mass $M_X$ to be much 
weaker than the numbers in Table~\ref{tab1}. For example,
this is the case if some of the elements ${\cal D}_{ed}$, ${\cal D}_{\mu d}$, 
${\cal D}_{es}$ and ${\cal D}_{\mu s}$ are small enough. 
Given the unitarity of the matrix ${\cal D}$, this would mean that 
its elements presented in the interactions of the $\tau$-lepton 
and $b$-quark should be close to unity.
In this case, leptoquarks might make more significant
contributions to $\tau$ and $B$ decays.

\begin{table}[t]
\tbl{Constraints on the model parameters from $\tau$ decays.}
{\begin{tabular}{lcl} \toprule
Experimental limit & Ref. & Bound \\ \colrule 
\bigskip
$Br(\tau^- \to e^- K^0_S ) < 2.6 \times 10^{-8}$ 
& \refcite{Miyazaki:2010} & $\frac{\mbox{\footnotesize $M_X$}}
{\mbox{\footnotesize 
$|{\cal D}_{e s}{\cal D}^*_{\tau d} - {\cal D}_{e d}{\cal D}^*_{\tau s}|^{1/2}$}} \,
> \, 10~\textrm{TeV}$ \\
\bigskip
$Br(\tau^- \to \mu^- K^0_S ) < 2.3 \times 10^{-8}$ 
& \refcite{Miyazaki:2010} & $\frac{\mbox{\footnotesize $M_X$}}
{\mbox{\footnotesize 
$|{\cal D}_{\mu s}{\cal D}^*_{\tau d} - {\cal D}_{\mu d}{\cal D}^*_{\tau s}|^{1/2}$}}
\,
> \, 11~\textrm{TeV}$ \\
\bigskip
$Br(\tau^- \to e^- \pi^0 ) < 8.0 \times 10^{-8}$ 
& \refcite{Miyazaki:2007} & $\frac{\mbox{\footnotesize $M_X$}}
{\mbox{\footnotesize $|{\cal D}_{ed} {\cal D}_{\tau d}|^{1/2}$}} \,
> \, 7~\textrm{TeV}$ \\
\bigskip
$Br(\tau^- \to \mu^- \pi^0 ) < 1.1 \times 10^{-7}$ 
& \refcite{Aubert:2007} & $\frac{\mbox{\footnotesize $M_X$}}
{\mbox{\footnotesize $|{\cal D}_{\mu d} {\cal D}_{\tau d}|^{1/2}$}}
\,
> \, 6~\textrm{TeV}$ \\
\botrule
\end{tabular}
\label{tab2}}
\end{table}

\subsection{$\tau$ Decays}

The current accuracy of experimental data on the $\tau$ and $B$ decays is 
poorer than in the processes considered above. Nevertheless it is possible
to obtain some constraints on the elements of the matrix ${\cal D}$ and on the
leptoquark mass from the data on decays that are strongly suppressed or forbidden 
in the standard model. 

The amplitude of the decay $\tau^-\to\mu^-K^0_S$ proceeding through a
leptoquark is given by 
\begin{equation} 
{\cal M}^X_{\tau\mu K} = 
\frac{\pi\al_S\l(M_X\r)f_K}{\sqrt{2} \,M_X^2} 
\left( {\cal D}_{\mu s}{\cal D}^*_{\tau d} - {\cal D}_{\mu d}{\cal D}^*_{\tau s} \right)
\l(m_\tau+m_\mu-\fc{2m_K^2Q}{m_s+m_d}\r) \l(\bar{\mu}\ga_5\tau\r).
\label{Amp5} 
\end{equation}

The amplitude of the decay $\tau^-\to e^-K^0_S$ is obtained from
\eqref{Amp5} as the result of the substitution $\mu\to e$. From the
experimental upper limits on the widths of these decays,\cite{Miyazaki:2010} 
we find the constraints which are presented in the Table~\ref{tab2}.

Making corresponding substitutions, one can obtain in a similar way
the constraints from the processes $\tau^-\to e^-\pi^0$ 
and $\tau^-\to\mu^-\pi^0$, see Table~\ref{tab2}. 

\subsection{Rare $B$-Meson Decays}

Let us consider the $B$ decays, for which there are experimental
constraints, and which may proceed through an intermediate
leptoquark. The decay $B^+\to K^+e^-\mu^+$ is an example of the process being 
forbidden in the standard model. The amplitude of this decay via the 
vector leptoquark can be represented in the form 
\begin{eqnarray} 
{\cal M}^X_{BKe\mu} &=& -\fc{2\st{2}\pi\al_S\l(M_X\r)}{M_X^2}{\cal D}_{es}{\cal D}^*_{\mu b} 
\bigg[\hf
f^0_+\l(q^2\r) \l(p_B+p_K\r)_\al\l(\bar{e}\ga_\al\mu\r)
\nonumber\\
&-&\fc{f^0_+\l(q^2\r)m_B^2+f^0_-\l(q^2\r)q^2}{m_b}Q\l(\mu_0\r)\l(\bar{e}\mu\r)\bigg],
\label{Amp6} 
\end{eqnarray} 
where $p_B$ and $p_K$ are the 4-momenta of the
$B$- and $K$-mesons; the remaining notation is identical to that in
equation \eqref{Amp4}. We assume that the form factors
$f^0_{+,-}$ are of the same order of magnitude as in the $K^0_{\ell 3}$
decay. This assumption is in agreement with the results obtained from
the analysis of the decay $B^+\to\overline{D}^0 \ell^+ \nu$. 

\begin{table}[t]
\tbl{Constraints on the model parameters from the data on rare $B^+$ decays.}
{\begin{tabular}{lcl} \toprule
Experimental limit & Ref. & Bound \\ \colrule 
\bigskip
$Br(B^+ \to K^+ e^- \mu^+) < 1.3 \times 10^{-7}$ 
& \refcite{Aubert:2006a} & $\frac{\mbox{\footnotesize $M_X$}}
{\mbox{\footnotesize $|{\cal D}_{es} {\cal D}_{\mu b}|^{1/2}$}} \,
> \, 38~\textrm{TeV}$ \\
\bigskip
$Br(B^+ \to K^+ e^+ \mu^-) < 0.91 \times 10^{-7}$ 
& \refcite{Aubert:2006a} & $\frac{\mbox{\footnotesize $M_X$}}
{\mbox{\footnotesize $|{\cal D}_{\mu s} {\cal D}_{eb}|^{1/2}$}} \,
> \, 42~\textrm{TeV}$ \\
\bigskip
$Br(B^+ \to \pi^+ e^+ \mu^-)<0.92\times10^{-7}$ 
& \refcite{Aubert:2007a} & $\frac{\mbox{\footnotesize $M_X$}}
{\mbox{\footnotesize $|{\cal D}_{\mu d} {\cal D}_{e b}|^{1/2}$}} \,
> \, 42~\textrm{TeV}$ \\
\bigskip
$Br(B^+ \to K^+ \mu^\pm \tau^\mp)<7.7\times10^{-5}$ 
& \refcite{Aubert:2007b} &  $\frac{\mbox{\footnotesize $M_X$}}
{\mbox{\footnotesize $\left(|{\cal D}_{\tau s} {\cal D}_{\mu b}|^2 
+|{\cal D}_{\mu s} {\cal D}_{\tau b}|^2 \right)^{1/4}$}} \,
> \, 8~\textrm{TeV}$ \\
\botrule
\end{tabular}
\label{tab3}}
\end{table}

The amplitude of the decay $B^+\to K^+e^+\mu^-$
is obtained from Eq.~\eqref{Amp6} by means of the interchange
$e\leftrightarrow\mu$. Using the 
experimental data reported in Ref.~\refcite{Aubert:2006a}, we arrive at the
constraints 
\begin{equation} 
\ba &M_X>\l(38~\text{TeV}\r)\l|{\cal D}_{es}{\cal D}_{\mu b}\r|^{1/2},\\
&M_X>\l(42~\text{TeV}\r)\l|{\cal D}_{\mu s}{\cal D}_{eb}\r|^{1/2}.
\label{Con10} \ea 
\end{equation} 
Making corresponding substitutions, one can also obtain in a similar
way as for the decays of the $K^+$ meson, the respective constraints 
from other processes of $B^+$ decay (see Table~\ref{tab3}).

The amplitudes of the $B^0$ and $B^0_s$ decay processes  
into charged lepton pairs can be calculated in a similar way 
as the amplitude \eqref{Amp3}. For the $B$ meson decay constant we take 
$f_B = 220~\textrm{MeV}$, see e.g. Refs.~\refcite{Jamin:2002,Ikado:2006}. 
From these processes the constraints on the leptoquark mass and the ${\cal D}$ matrix elements 
are obtained. They are collected in Table~\ref{tab4}.

\begin{table}[t]
\tbl{Constraints on the model parameters from 
experimental data on rare $B^0$ and $B^0_s$ decays.}
{\begin{tabular}{lcl} \toprule
Experimental limit & Ref. & Bound \\ \colrule 
\bigskip
$Br(B^0\to e^+e^-)<8.3\times10^{-8}$ 
& \refcite{Aaltonen:2009} & $\frac{\mbox{\footnotesize $M_X$}}
{\mbox{\footnotesize $|{\cal D}_{ed} {\cal D}_{eb}|^{1/2}$}} \,
> \, 51~\textrm{TeV}$ \\
\bigskip
$Br(B^0\to\mu^+\mu^-)<1.5\times10^{-8}$ 
& \refcite{Aaltonen:2008} & $\frac{\mbox{\footnotesize $M_X$}}
{\mbox{\footnotesize $|{\cal D}_{\mu d} {\cal D}_{\mu b}|^{1/2}$}} \,
> \, 79~\textrm{TeV}$ \\
\bigskip
$Br(B^0\to\tau^+\tau^-)<4.1\times10^{-3}$ 
& \refcite{Aubert:2006b} & $\frac{\mbox{\footnotesize $M_X$}}
{\mbox{\footnotesize $|{\cal D}_{\tau d} {\cal D}_{\tau b}|^{1/2}$}} \,
> \, 3~\textrm{TeV}$ \\
\bigskip
$Br(B^0\to e^+ \mu^-)<6.4\times10^{-8}$ 
& \refcite{Aaltonen:2009} 
& $\frac{\mbox{\footnotesize $M_X$}}
{\mbox{\footnotesize $|{\cal D}_{\mu d} {\cal D}_{e b}|^{1/2}$}} \,
> \, 55~\textrm{TeV}$ \\
\bigskip
$Br(B^0\to e^+ \tau^-)<2.8\times10^{-5}$ 
& \refcite{Aubert:2008} 
& $\frac{\mbox{\footnotesize $M_X$}}
{\mbox{\footnotesize $|{\cal D}_{\tau d} {\cal D}_{e b}|^{1/2}$}} \,
> \, 11~\textrm{TeV}$ \\
\bigskip
$Br(B^0\to \mu^+ \tau^-)<2.2\times10^{-5}$ 
& \refcite{Aubert:2008} 
& $\frac{\mbox{\footnotesize $M_X$}}
{\mbox{\footnotesize $|{\cal D}_{\tau d} {\cal D}_{\mu b}|^{1/2}$}} \,
> \, 12~\textrm{TeV}$ \\
\bigskip
$Br(B^0_s\to e^+e^-)<2.8\times10^{-7}$ 
& \refcite{Aaltonen:2009} & $\frac{\mbox{\footnotesize $M_X$}}
{\mbox{\footnotesize $|{\cal D}_{es} {\cal D}_{eb}|^{1/2}$}} \,
> \, 38~\textrm{TeV}$ \\
\bigskip
$Br(B^0_s\to\mu^+\mu^-)<4.2\times10^{-8}$ 
& \refcite{Abazov:2010} & $\frac{\mbox{\footnotesize $M_X$}}
{\mbox{\footnotesize $|{\cal D}_{\mu s} {\cal D}_{\mu b}|^{1/2}$}} \,
> \, 61~\textrm{TeV}$ \\
\bigskip
$Br(B^0_s\to e^+ \mu^-)<2.0\times10^{-7}$ 
& \refcite{Aaltonen:2009} & $\frac{\mbox{\footnotesize $M_X$}}
{\mbox{\footnotesize $|{\cal D}_{\mu s} {\cal D}_{e b}|^{1/2}$}} \,
> \, 41~\textrm{TeV}$ \\
%
\botrule
\end{tabular}
\label{tab4}}
\end{table}

\section{Combined constraint from the $\pi$, $K$, $\tau$, $B$ decays}
\label{sec:Combination}

All the constraints on the vector leptoquark mass collected in Tables~\ref{tab1}--\ref{tab4} 
contain the elements of the unitary ${\cal D}$ matrix:
\begin{equation} 
{\cal D}_{\ell n} = 
\begin{pmatrix} 
{\cal D}_{e d} & {\cal D}_{e s} & {\cal D}_{e b}\\[2mm]
{\cal D}_{\mu d} & {\cal D}_{\mu s} & {\cal D}_{\mu b}\\[2mm]
{\cal D}_{\tau d} & {\cal D}_{\tau s} & {\cal D}_{\tau b}
\end{pmatrix}.
\label{Ddef} 
\end{equation}
There are also two constraints in Table~\ref{tab1} 
containing the element of the ${\cal U}$ matrix: 
\begin{equation} 
{\cal U}_{e u} = 
{\cal D}_{e d} V^*_{u d} + {\cal D}_{e s} V^*_{u s} + {\cal D}_{e b} V^*_{u b} \simeq 
0.974 \,{\cal D}_{e d} + 0.225 \,{\cal D}_{e s} + 0.004 \,{\cal D}_{e b}  
\,,
\label{U_eu} 
\end{equation}
where $V$ is the Cabibbo--Kobayashi--Maskawa quark-mixing matrix, 
see Ref.~\refcite{Nakamura:2010}, p.~150.

Let us try to establish, varying the unknown elements of the ${\cal D}$ matrix, 
the lowest limit on the vector leptoquark mass, being in agreement with 
all the constraints presented in Tables~\ref{tab1}--\ref{tab4}. 
The similar approach was developed in Ref.~\refcite{Povarov:2011}. 

It is possible for the strongest constraint on $M_X$ arising from the limit on 
$Br(K^0_L \to e^\pm \mu^\mp)$, see Table~\ref{tab1}, to be much lower than 2100 TeV if the matrix elements 
${\cal D}_{e d}$ and ${\cal D}_{e s}$ are small enough. For evaluation, let us 
take them to be zero. In this case, all the estimates which are presented 
in the right column of Table~\ref{tab1}, disappear,
except the one arising from the limit on $Br(K^0_L \to \mu^+ \mu^-)$. 
Because of unitarity of the ${\cal D}$ matrix, the elements ${\cal D}_{\mu b}$ and 
${\cal D}_{\tau b}$ are also equal to zero. 
The remaining $(2 \times 2)$-matrix can be parameterized by one angle. 
The insertion of the phase factor allows to eliminate the restriction arising from the  
limit on $Br(K^0_L \to \mu^+ \mu^-)$ which contains the real part of the 
${\cal D}$ matrix elements product. For example, it is possible to take the ${\cal D}$ matrix 
in the form:
\begin{equation} 
{\cal D}_{\ell n} \simeq 
\begin{pmatrix} 
0 & 0 & 1~\\[2mm]
\cos \varphi & ~\text{i} \sin \varphi~  & 0~\\[2mm]
~\text{i} \sin \varphi & \cos \varphi & 0~
\end{pmatrix}.
\label{Dfin} 
\end{equation}
As the analysis shows, in this case there appear the following constraints 
from the remaining limits of Tables~\ref{tab2}--\ref{tab4} 
on the branching ratios of the processes:
\newline
i) $\tau^- \to \mu^- K^0_S$
\begin{equation} 
M_X > 11~\textrm{TeV} \, |\cos 2 \varphi|^{1/2} \,,
\label{fin1} 
\end{equation}
ii) $\tau^- \to \mu^- \pi^0$
\begin{equation} 
M_X > 6~\textrm{TeV} \, |\sin 2 \varphi|^{1/2} \,,
\label{fin2} 
\end{equation}
iii) $B^0\to e^+ \mu^-$
\begin{equation} 
M_X > 55~\textrm{TeV} \, |\cos \varphi|^{1/2} \,,
\label{fin3} 
\end{equation}
iv) $B^0_s\to e^+ \mu^-$
\begin{equation} 
M_X > 41~\textrm{TeV} \, |\sin \varphi|^{1/2} \,.
\label{fin4} 
\end{equation}
Here, the weaker constraints of the same type are omitted. 
Combining these constraints one obtains the final limit on the vector leptoquark mass 
from low-energy processes: 
\begin{equation} 
M_X > 38~\textrm{TeV} \,.
\label{finMX} 
\end{equation}
%
\section{Different mixings for left-handed and right-handed fermions}
\label{sec:Different}

\begin{table}[t]
\tbl{Constraints on the model parameters from 
experimental data on rare $\pi$ and $K$ decays in the case of 
different mixings for left-handed and right-handed fermions.}
{\begin{tabular}{ll} \toprule
Experimental data & Bound \\ \colrule 
\bigskip
$\frac{\mbox{\footnotesize $\Gamma\left(\pi^+ \to e^+ \nu_e\right)$}}
{\mbox{\footnotesize $\Gamma\left(\pi^+ \to \mu^+ \nu_\mu\right)$}}$ 
& $\frac{\mbox{\footnotesize
$M_X$}} {\mbox{\footnotesize $\left|\text{Re}({\cal D}^{(R)}_{ed} {\cal U}^{(L)*}_{eu}) \right|^{1/2}$}} \,
> \, 210~\textrm{TeV}$ \\
\bigskip
$\frac{\mbox{\footnotesize $\Gamma\left(K^+ \to e^+ \nu_e\right)$}}
{\mbox{\footnotesize $\Gamma\left(K^+ \to \mu^+ \nu_\mu\right)$}}$ 
& $\frac{\mbox{\footnotesize $M_X$}}
{\mbox{\footnotesize $\left|\text{Re}({\cal D}^{(R)}_{es} {\cal U}^{(L)*}_{eu})\right|^{1/2}$}} \,
> \, 150~\textrm{TeV}$ \\
\bigskip
$Br(K^+ \to \pi^+ \mu^+ e^-)$ & $\frac{\mbox{\footnotesize $M_X$}}
{\mbox{\footnotesize 
$\left(\left|{\cal D}^{(L)}_{ed} {\cal D}^{(R)}_{\mu s}\right|^2 
+ \left|{\cal D}^{(R)}_{ed} {\cal D}^{(L)}_{\mu s}\right|^2 
\right)^{1/4}$}} \,
> \, 200~\textrm{TeV}$ \\
\bigskip
$Br(K^+ \to \pi^+ \mu^- e^+)$ & $\frac{\mbox{\footnotesize $M_X$}}
{\mbox{\footnotesize 
$\left(\left|{\cal D}^{(L)}_{es} {\cal D}^{(R)}_{\mu d}\right|^2 
+ \left|{\cal D}^{(R)}_{es} {\cal D}^{(L)}_{\mu d}\right|^2 
\right)^{1/4}$}} \,
> \, 84~\textrm{TeV}$ \\
\bigskip
$Br(K^0_L \to \mu^+ \mu^-)$ & $\frac{\mbox{\footnotesize
$M_X$}} {\mbox{\footnotesize 
$\left|{\cal D}^{(L)}_{\mu d} {\cal D}^{(R)*}_{\mu s} 
+ {\cal D}^{(R)*}_{\mu d} {\cal D}^{(L)}_{\mu s}\right|^{1/2}$}} \,
> \, 780~\textrm{TeV}$ \\
\bigskip
$Br(K^0_L \to e^\pm \mu^\mp)$ & $\frac{\mbox{\footnotesize $M_X$}}
{\mbox{\footnotesize 
$\left(\left|{\cal D}^{(L)}_{ed} {\cal D}^{(R)*}_{\mu s}+{\cal D}^{(L)}_{es} {\cal D}^{(R)*}_{\mu d}\right|^2 
+ \{ L \leftrightarrow R \}
\right)^{1/4}$}} \,
> \, 1770~\textrm{TeV}$ \\
\bigskip
$Br(K_L^0 \to e^+ e^-)$ & $\frac{\mbox{\footnotesize $M_X$}}
{\mbox{\footnotesize 
$\left|{\cal D}^{(L)}_{e d} {\cal D}^{(R)*}_{e s} 
+ {\cal D}^{(R)*}_{e d} {\cal D}^{(L)}_{e s}\right|^{1/2}$}} \,
> \, 1700~\textrm{TeV}$ \\
\botrule
\end{tabular}
\label{tab5}}
\end{table}

We consider a possibility when the quark-lepton symmetry 
is the next step beyond the standard model. Then the left-right symmetry which is 
believed to exist in Nature, should restore at higher mass scale. 
But this means that the left-right symmetry should be broken at the scale $M_X$. 
It is worthwhile to consider the matrices ${\cal D}^{(L)}, {\cal U}^{(L)}$ and 
${\cal D}^{(R)}, {\cal U}^{(R)}$ which are in a general case different 
for left-handed and right-handed fermions. 
This possibility and some its consequences were also considered 
in Refs.~\refcite{Smirnov:1995a}--\refcite{Smirnov:2008}.
The interaction Lagrangian of leptoquarks with fermions takes the form instead of 
Eq.~\eqref{LagDen}:
\begin{eqnarray}
{\cal L}_X &=& \fc{g_S\l(M_X\r)}{2 \st{2}} \bigg[ 
{\cal D}^{(L)}_{\ell n} \left(\bar{\ell} O_\al d_n \right) + 
{\cal D}^{(R)}_{\ell n} \left(\bar{\ell} O_\al^\prime d_n \right) 
\nonumber\\
&+& \left( {\cal K}^{(L)\dagger} {\cal U}^{(L)} \right)_{ip} \l(\bar{\nu}_i O_\al u_p \right) 
+ \left( {\cal K}^{(R)\dagger} {\cal U}^{(R)} \right)_{ip} \l(\bar{\nu}_i O_\al^\prime u_p \right) 
\bigg] 
X_\al + \text{h.c.},
\label{Lagr_LR} 
\end{eqnarray} 
where $O_\al=\ga_\al\l(1-\ga_5\r)$, $O_\al^\prime=\ga_\al\l(1+\ga_5\r)$. 

The constraints on the model parameters 
from experimental data on rare $\pi$ and $K$ decays collected in Table~\ref{tab1} 
in the case of different mixings take the forms presented in Table~\ref{tab5}. 
If one would wish to reduce the limits of Table~\ref{tab5} on $M_X$ from thousands and hundreds 
to tens of TeV by varying the elements of the ${\cal D}^{(L)}$ and ${\cal D}^{(R)}$ matrices, it seems that the elements ${\cal D}^{(L)}_{e d}$ and ${\cal D}^{(R)}_{e d}$ should be taken small in any case.  
For evaluation, let them be zero. Then, the most strong restriction of Table~\ref{tab5} from 
the limit on $Br(K^0_L \to e^\pm \mu^\mp)$ takes the form:
\begin{equation} 
\frac{M_X}
{\left(\left|{\cal D}^{(L)}_{es} {\cal D}^{(R)}_{\mu d}\right|^2 
+ \left|{\cal D}^{(R)}_{es} {\cal D}^{(L)}_{\mu d}\right|^2 
\right)^{1/4}} > 1770~\textrm{TeV} \,.
\label{LR_2} 
\end{equation}

There are two possibilities to eliminate this bound together with other bounds 
of Table~\ref{tab5} which we call the symmetric and the asymmetric cases.

\emph{The symmetric case} is realized when both of the matrices ${\cal D}^{(L)}$ 
and ${\cal D}^{(R)}$ are taken in the form of Eq.~\eqref{Dfin} with the angles $\varphi_L$ and $\varphi_R$. 
In this case the restriction from the limit on $Br(K^0_L \to \mu^+ \mu^-)$ 
takes the form:
\begin{equation} 
M_X > 780~\textrm{TeV} \, |\sin \left( \varphi_L - \varphi_R \right)|^{1/2} \,,
\label{LR_3} 
\end{equation}
and the angles should be close to each other or differ by $\pi$, in any case 
we come back to the results of Sec.~\ref{sec:Combination}. 

\emph{The asymmetric case} is realized when the matrices are taken in the form:
\begin{equation} 
{\cal D}^{(L)}_{\ell n} \simeq 
\begin{pmatrix} 
~0 & \cos \chi_L &  ~\sin \chi_L~\\[2mm]
~0 & - \sin \chi_L & ~\cos \chi_L~\\[2mm]
~1 & 0 & 0
\end{pmatrix} ,
\quad
{\cal D}^{(R)}_{\ell n} \simeq 
\begin{pmatrix} 
~0~~ & 0~~ & 1~~\\[2mm]
~0~~ & 1~~ & 0~~\\[2mm]
~1~~ & 0~~ & 0~~
\end{pmatrix} .
\label{LR_4} 
\end{equation}
In the Table~\ref{tab4} which has provided one more group of essential restrictions 
in Sec.~\ref{sec:Combination}, the following substitutions should be made 
in the bounds:
\begin{equation} 
|{\cal D}_{\ell q} {\cal D}_{\ell^\prm b}| \Rightarrow
\frac{1}{\sqrt{2}} \left( \left|{\cal D}^{(L)}_{\ell q} {\cal D}^{(R)}_{\ell^\prm b} \right|^2 +
\left|{\cal D}^{(R)}_{\ell q} {\cal D}^{(L)}_{\ell^\prm b} \right|^2 \right)^{1/2} ,
\label{LR_5} 
\end{equation}
where $\ell, \ell^\prm = e, \mu, \tau$ and $q = d, s$.
As the analysis shows, the most stringent constraints arise 
from the following limits on the branching ratios of the processes:
\newline
i) $B^0_s \to \mu^+ \mu^-$
\begin{equation} 
M_X > 51~\textrm{TeV} \, |\cos \chi_L|^{1/2} \,,
\label{LR_6} 
\end{equation}
ii) $B^0_s \to e^+ \mu^-$
\begin{equation} 
M_X > 41~\textrm{TeV} \, |\sin \chi_L|^{1/2} \,.
\label{LR_7} 
\end{equation}
Combining these constraints, one obtains the final limit on the vector leptoquark mass 
from low-energy processes in the case of different mixing matrices for left-handed and right-handed 
fermions, which coincides, with a good accuracy, with the limit \eqref{finMX} obtained 
in the left-right-symmetric case: 
\begin{equation} 
M_X > 38~\textrm{TeV} \,.
\label{LR_8} 
\end{equation}
%

\section{Constraint from the decay $\pi^0\to\nu\bar{\nu}$}
\label{sec:Pi_decay}

Only one process was found\cite{Kuznetsov:1994} in which the lower limit on the
leptoquark mass was independent of mixing parameters, the decay
$\pi^0\to\nu\bar{\nu}$. 

In the standard model, the width of this process is proportional to
$m_\nu^2$, but it can also proceed through the leptoquark exchange;
in the latter case, there is no suppression associated with the
smallness of the neutrino mass. The corresponding amplitude of the
process has the form 
\begin{equation} {\cal
M}^X_{\pi\nu\nu}=\fc{\pi\al_S\l(M_X\r)f_\pi
m_\pi^2Q}{\st{2}M_X^2m_u}\l({\cal K}^\dagger {\cal U}\r)_{iu} 
\l({\cal U}^\dagger {\cal K}\r)_{uj}\l(\bar{\nu}_i\ga_5\nu_j\r), 
\label{Amp7} 
\end{equation} 
and the decay probability summed over all neutrino flavors $i$ and $j$
does not depend on mixing. 

From the accelerator data\cite{Artamonov:2005} for the decay $\pi^0\to\nu\bar{\nu}$:
\begin{equation}
\text{Br}\l(\pi^0\to\nu\bar{\nu}\r)<2.7\times10^{-7}, 
\label{BrPi02} 
\end{equation}
we obtain a constraint: 
\begin{equation} 
M_X>600~\text{GeV}. 
\label{Con12} 
\end{equation}

In Refs.~\refcite{Natale:1991,Lam:1991} the almost coinciding astrophysical and cosmological estimations 
of the width of this decay were found, being much stronger than the accelerator limit~\eqref{BrPi02}, 
\begin{equation}
\text{Br}\l(\pi^0\to\nu\bar{\nu}\r)<3\times10^{-13}. 
\label{BrPi01} 
\end{equation}
The resulting constraint on the leptoquark mass was estimated to be:\cite{Kuznetsov:1994} 
\begin{equation} 
M_X>18~\text{TeV}. 
\label{Con11} 
\end{equation}

The astrophysical estimation\cite{Natale:1991} was based on evaluating the excess energy-loss rate 
from SN 1987A if the process $\gamma\gamma\to\pi^0\to\nu\bar{\nu}$ via the pion-pole mechanism occurred, 
permitted if the neutrinos had a right-handed component.
In turn, the cosmological limit on the width of the decay $\pi^0\to\nu\bar{\nu}$ 
was established in Ref.~\refcite{Lam:1991}, where 
the production was considered of right-handed neutrinos produced from the cosmic
thermal background at the temperature of about the pion mass through the reaction
$\gamma\gamma\to\pi^0\to\nu\bar{\nu}$. 

However, in Ref.~\refcite{Raffelt:1991} it was mentioned that the astrophysical 
limit\cite{Natale:1991} must be relaxed in several orders of magnitude, because 
the effect of nuclear absorption of a pion in a supernova core was not considered 
in Ref.~\refcite{Natale:1991}. In turn, a criticism has been expressed 
in Ref.~\refcite{Gregores:1995} on the cosmological limit\cite{Lam:1991} also. 
At the temperature where the pion mechanism is at resonance, the strong interaction 
among pions occurs much faster than the pion decay: the rate of $\pi - \pi$
scattering, $\Gamma_{\pi - \pi} \sim$ 0.2 MeV, dominates the pion lifetime
in the dense medium, resulting in a suppression of several
orders of magnitude in the rate of neutrino production.

Therefore, only the laboratory limit\cite{Artamonov:2005} for the decay $\pi^0\to\nu\bar{\nu}$
should be considered as reliable, to establish the bound $M_X>0.6~\text{TeV}$. 

\section{Conclusion}

Thus, the detailed analysis of the available experimental data on rare
$\pi$, $K$, $\tau$ and $B$ decays and on the $\mu e$ conversion yields constraints on
the vector leptoquark mass that always involve the elements of the
unknown mixing matrix ${\cal D}$. 
Combining the constraints from the experimental data on the low-energy processes 
presented in Tables~\ref{tab1}--\ref{tab4}, 
we have obtained in the case of identical mixings for left-handed and right-handed fermions
the following lowest limit on the vector leptoquark mass:  
$M_X > 38~\textrm{TeV}$. The lowest limit obtained in the case of different mixing matrices 
for left-handed and right-handed fermions appears to be the same. 

\section*{Acknowledgments}

We thank A.\,V.~Povarov and A.\,D.~Smirnov for useful discussions. 

The study was performed within the State Assignment for Yaroslavl 
University (Project \No~2.4176.2011), and was supported in part by the 
Russian Foundation for Basic Research (Project \No~11-02-00394-a).


\end{document}